\documentstyle[12pt]{article}
\def\bold#1{\setbox0=\hbox{$#1$}%
      \kern-.025em\copy0\kern-\wd0
      \kern.05em\copy0\kern-\wd0
      \kern-.025em\raise.0433em\box0 }

\def\eea{\end{eqnarray}}
\def\bea{\begin{eqnarray}}
\def\eeas{\end{eqnarray*}}
\def\beas{\begin{eqnarray*}}
\def\ee{\end{equation}}
\def\be{\begin{equation}}
\def\bdm{\begin{displaymath}}
\def\edm{\end{displaymath}}

\def\Tr{\mbox{Tr}}
\def\skp{\epsilon}

\renewcommand{\thefootnote}{\fnsymbol{footnote}}
\begin{document}
\begin{titlepage}


 
\begin{center}

\vspace*{2.0cm}
{\large\bf SU(3) SYMMETRY BREAKING AND OCTET BARYON 
POLARIZABILITIES\footnote[1]{Supported in part by the 
Deutsche Forschungsgemeinschaft (DFG) under contract 
Re 856/2--2.}}
\vskip 1.0cm

{Norberto N. SCOCCOLA$^{a,b}$\footnote[2]{Fellow
of the CONICET, Argentina.}, Herbert WEIGEL$^c$ 
and Bernd SCHWESINGER$^d$}
\vskip .2cm
{\it
$^a$ Physics Department, Comisi\'on Nacional de Energ\'{\i}a At\'omica,
          Av.Libertador 8250, (1429) Buenos Aires, Argentina. \\
$^b$ INFN, Sezione di Milano, via Celoria 16, I--20133 Milano, Italy.\\  
$^c$ Institute for Theoretical Physics, T\"ubingen University,
     Auf der Morgenstelle 14, D--72076 T\"ubingen, Germany.\\
$^d$ Siegen University, Fachbereich Physik, 
     D--57068 Siegen, Germany. \\}

\vskip 4.cm
{\bf ABSTRACT}\\
\begin{quotation}
Static polarizabilities of the low--lying $\frac{1}{2}^+$ 
baryons are studied within the collective coordinate approach to 
the three flavor generalization of the Skyrme model; in particular, 
magnetic polarizabilities are considered.
Predicted polarizabilities, which result from different
treatments of the strange degrees of freedom in this model, are 
critically compared. Their deviations from the 
flavor symmetric formulations are discussed.
\end{quotation}
\end{center}
\vskip 0.5cm
\leftline{\it PACS: 12.39.Dc, 14.20.Jn}
\vskip 0.5cm
\leftline{\it Keywords: \parbox[t]{15cm}{Hyperon polarizabilities,
Flavor symmetry breaking, Skyrmion,\\ Collective coordinates}}
\end{titlepage}

\renewcommand{\thefootnote}{\arabic{footnote}}

At Fermilab the $\Sigma$ hyperon polarizabilities will soon be
measured \cite{FNAL94,Moi94} and hyperon beams at CERN will provide 
data on the polarizabilities of other hyperons as well. In addition,
a rather precise determination of the nucleon polarizabilities 
is available \cite{Mac95}. This is of great interest because the
electromagnetic polarizabilities contain important information on
the baryon structure\cite{Pet81}. Although a rather large number 
of theoretical work has been devoted to the nucleon electromagnetic 
polarizabilities (see Ref.\cite{Lvo93} for a recent review) 
only quite recently the hyperon polarizabilities have been 
investigated. In Ref.\cite{LM92} the electric and magnetic
polarizabilities of the $\Sigma_+$ and $\Sigma_-$ hyperons were
computed within the non-relativistic quark model. A study of the
hyperon polarizabilities in heavy baryon chiral perturbation theory
has been reported in Ref.\cite{BKKM92}. Within the chiral soliton
models, predictions for hyperon electric polarizabilities using
the $SU(3)$ collective coordinate approach have been given
in Ref.\cite{Sch93}. Results 
for electric and magnetic static polarizabilities obtained
within an alternative treatment
of strange mesons in soliton models, the so-called bound state 
approach (BSA), \cite{GSS96} , have
been given recently. In this context the purpose of the present 
work is twofold. Firstly, we will continue the study of the 
polarizabilities in the $SU(3)$ collective coordinate approach
to the soliton model by presenting the corresponding predictions
for the static magnetic polarizabilities. Secondly, we will critically
analyze and compare the results obtained within the different 
approaches to baryons within the $SU(3)$ Skyrme 
model.

Our starting point is a gauged effective chiral action
\be
\Gamma = 
\int d^4 x \Big\{ {f^2_\pi \over 4}
\Tr\left[ D_\mu U (D^\mu U)^\dagger \right]
+
 {1\over{32 \skp^2}}
 \Tr\left[ [U^\dagger D_\mu U , U^\dagger D_\nu U]^2\right] \Big\}
+ \Gamma_{an} + \Gamma_{sb}
\label{lag}
\ee
Here $f_\pi=93{\rm MeV}$ is the pion decay constant and $\epsilon$ 
is the dimensionless Skyrme parameter. Furthermore the chiral field 
$U$ is the non--linear realization of the pseudoscalar octet. 
The covariant derivative is defined as
\be
D_\mu U = \partial_\mu U + ie\ A_\mu \ [Q, U ] \ ,
\qquad 
Q = {1\over2} \ \left[ \lambda_3 + {1\over{\sqrt{3}}} \lambda_8
\right] \, ,
\ee
where $A_\mu$ is the electromagnetic field and $Q$ the electric 
charge matrix. Throughout this paper we adopt Gaussian units,
{\it i.e.} $e^2=1/137$. In Eq (\ref{lag}) $\Gamma_{an}$ is the 
Wess-Zumino action gauged to contain the electromagnetic interaction 
\cite{Wit83} while the (gauged) symmetry breaking term $\Gamma_{sb}$ 
\cite{GSS96} accounts for different masses and decay constants 
of the pseudoscalar fields \cite{We90}.
It is convenient to order the 
effective action according to powers of $A_\mu$
\be \label{gafin}
\Gamma = \Gamma^{strong} +
        \Gamma^{lin} + \Gamma^{quad} \, .
\ee
Formally we may write
\bea
\Gamma^{lin}  =  \int d^4x \ e \ A_\mu J^\mu \, 
\quad {\rm and} \quad
\Gamma^{quad}  =  -\int d^4x \ e^2 \ A_\mu \ G^{\mu\nu} \ A_\nu \, .
\eea
Explicit expressions for the electromagnetic current $J^\mu$ and 
the seagull tensor $G^{\mu\nu}$ can {\it e.g.} be found in 
ref \cite{GSS96}. Actually both $\Gamma^{lin}$ and $\Gamma^{quad}$ 
contribute to the baryon polarizabilities. In second order 
perturbation $\Gamma^{lin}$ gives rise to the so-called ``dispersive"
contributions while $\Gamma^{quad}$ yields the so-called
``seagull" contributions.

In Eq.(\ref{gafin}) $\Gamma^{strong}$ is the action in the absence of 
the electromagnetic field.  In the soliton picture strong interaction 
properties of the low--lying $\frac{1}{2}^+$ and $\frac{3}{2}^+$ 
baryons are computed following the standard $SU(3)$ collective 
coordinate approach to the Skyrme model. We introduce the 
{\it ansatz} 
\begin{equation}
U({\bf r}, t) = A(t) \ \left( 
\begin{array}{cc} 
c + i \mbox{\boldmath $\tau$} \cdot 
{\hat{\mbox{\boldmath $r$}}} \ s & 0 \\
0  & 1 
\end{array}
\right) 
\ A^\dagger(t) \ 
\label{ansatz}
\end{equation}
for the chiral field.
Here we have employed the abbreviations $c= \cos F(r)$ and 
$s=\sin F(r)$ where $F(r)$ is the chiral angle which parametrizes 
the soliton. The collective rotation matrix $A(t)$ is $SU(3)$ valued. 
Substituting the configuration (\ref{ansatz}) into $\Gamma^{strong}$
yields (upon canonical quantization of $A$) the collective Hamiltonian.
Its eigenfunctions and eigenvalues are identified as the baryon 
wavefunctions $\Psi_B(A)=\langle B |A\rangle$ and masses $m_B$.
Due the symmetry breaking terms in $\Gamma_{sb}$ this Hamiltonian 
is obviously not $SU(3)$ symmetric. As shown by Yabu and Ando \cite{YA88}
it can, however, be diagonalized exactly. This diagonalization 
essentially amounts to admixtures of states from higher dimensional 
$SU(3)$ representations into the octet ($J=\frac{1}{2}$) and decouplet 
($J=\frac{3}{2}$) states. This procedure, commonly
known as ``Rigid Rotator Approach" (RRA), has proven quite successful 
in describing the hyperon spectrum and static properties \cite{We96}. 
In ref \cite{SW92} the chiral 
angle was allowed to adjust itself according the flavor orientation $A$. This 
approach considers the collective rotation as slow enough to let the 
soliton profile react on the forces exerted by the symmetry breaking, 
hence the notion ``Slow Rotator Approach" (SRA). In the SRA the
chiral angle not only depends on the radial coordinate $r$ but also
parametrically on the flavor orientation $A$. In contrast to both 
the RRA as well as the BSA this approach has the 
desired feature that the meson profiles of the configuration which 
have their chiral field rotated maximally into the strange direction 
decay with the kaon mass. The comparison \cite{SW92} of the 
predicted magnetic moments with the experimental data shows that the 
incorporation of symmetry breaking effects into the chiral angle is 
crucial to properly describe the observed deviations from $U$-spin 
symmetry\footnote{Similar results have been found by treating the 
influence of the symmetry breaking on the soliton extension at the 
quantum level \cite{Sch91}.}. It is a major purpose of the present 
paper to compare the predictions for the magnetic polarizabilities 
in these approaches to the three flavor Skyrme model.

The static polarizabilities can be extracted from the shift of
the particle energies in the presence of constant external 
electric ($\mbox{\boldmath $E$}$) and magnetic 
($\mbox{\boldmath $B$}$) fields:
\bea
\delta M &=& - {1\over2} \ \alpha \ \mbox{\boldmath $E$}^2
- {1\over2} \ \beta \ \mbox{\boldmath $B$}^2 \label{shift}  \, .
\eea
The electric ($\alpha$) and magnetic ($\beta$) polarizabilities
characterize the dynamical response to the external electromagnetic 
fields. Here we will concentrate on the magnetic polarizability 
$\beta$, which is easily obtained from 
(\ref{gafin}) by adopting 
\be
A^\mu = (0, -{1 \over 2} \bold r \times \bold B)\ .
\ee
In analogy to Eq. (\ref{gafin}) the Hamiltonian is expanded up 
to quadratic order in $\mbox{\boldmath $B$}$ 
\be \label{H}
H = H^{strong} +
        H^{lin} + H^{quad} \, .
\ee
The quadratic part yields the seagull contribution
$\beta_s$. Using the ansatz Eq.(\ref{ansatz}) one obtains
for $\frac{1}{2}^+$ baryons 
\begin{eqnarray} 
\beta^B_s &=&  \langle B \vert \left[
\gamma_\pi^{(m)}  D_{e,i}^2 + \gamma_K^{(m)} D_{e,\alpha}^2
\right] 
\vert B \rangle \ .
\label{betas}
\end{eqnarray}
These matrix elements are understood in the space 
of the collective coordinates with $D_{a,b}=\frac{1}{2}{\rm Tr}
\left(\lambda_a A \lambda_b A^\dagger\right)$ denoting the 
adjoint representation of the collective rotations. We have 
used the notation $i=1,2,3$ and $\alpha=4,5,6,7$. 
Moreover, a sum over repeated indices is understood and 
$D_{e,a} = D_{3,a} + {1\over{\sqrt3}} \ D_{8,a}$ refers to 
the electromagnetic direction. As discussed above, in the SRA the 
chiral angle depends on the flavor orientation $A$. Hence the 
spatial integrals 
\begin{eqnarray}
\gamma_\pi^{(m)} &=& - {e^2\over9} \int d^3 r \ r^2 s^2 
 \left[ f_\pi^2 + {1\over{\epsilon^2}} (F'^2 + {s^2\over{2 r^2}}) +
 {2\over3} \ (f_K^2 - f_\pi^2)\ c \ \left( 1- D_{8,8} \right) \right]
\hskip1.5cm~
\\
\gamma_K^{(m)} &=& - {e^2\over{12}} \int d^3 r \ r^2 (1-c)
\left[ f_K^2 + {1\over{4\epsilon^2}} (F'^2 + {s^2\over{r^2}}) +
(f_K^2 - f_\pi^2) { c - 2 \over3} 
\left( 1- D_{8,8} \right) \right] 
\end{eqnarray}
have both explicit and implicit dependencies on $A$. This has to 
be taken care of when computing the matrix elements (\ref{betas}) in 
the SRA.

The dispersive contribution $\beta_d$ arises from $H^{lin}$ in (\ref{H}).
Choosing the $z$--axis along the $\mbox{\boldmath $B$}$ field yields
in second order perturbation 
\be
\beta_d^B =  {e^2\over{2 M_N^2}} \sum_{B'\neq B} {{\vert \langle B \vert
\mu_3 \vert B' \rangle \vert^2   }\over {m_{B'} - m_{B}} }\ .
\label{betad}
\ee
Here $B$ and $B'$ refer to different baryon states and $\mu_3$ is 
the magnetic moment operator.  Its explicit expression for the present 
model can {\it e.g.} be found in Eqs.(13,15) of Ref.\cite{SW92}. In 
order to compute the dispersive magnetic polarizability of a given 
baryon $B$ we have to consider all possible states which are 
accessible from $B$ by magnetic dipole transitions. The dominant
contributions are expected from the lowest states with 
$|\triangle J|=|J-J^\prime|=1$ as these not only have the smallest 
mass differences but also sizable isovector contribution to the
magnetic transitions \cite{Sch95}. For example, in the case of the 
nucleon the $N\Delta$ transition would then be dominant. In addition, 
on top of the ground state in a given spin--isospin channel the 
$SU(3)$ collective coordinate approach predicts states, which have 
their major support from higher dimensional representations of 
$SU(3)$. For example, states with proton quantum numbers also exist in 
the ${\overline{\mbox{\boldmath $10$}}}$ and $\mbox{\boldmath $27$}$ 
representations. Such states also have non--vanishing magnetic 
dipole transitions to $B$. In Eq. (\ref{betad}) we have therefore 
included the magnetic dipole transitions to these states in both 
the rigid and the slow rotator approaches. Only in the limit of 
infinitely large symmetry breaking, when the model essentially 
reduces to flavor $SU(2)$, these transitions vanish.

The use of $f_K \ne f_\pi$ is essential to reproduce the experimentally
observed mass differences of the low--lying $\frac{1}{2}^+$ 
and $\frac{3}{2}^+$ baryons \cite{We90}. For definiteness we will 
take $f_K=120 MeV$ and $\epsilon =4.10$ \cite{PSW91} for RRA
and $f_K=118 MeV$ and $\epsilon=3.46$ \cite{SW92} for SRA respectively. For 
the meson masses we employ $m_\pi=138{\rm MeV}$ and $m_K=495{\rm MeV}$ 
in both cases. In tables 1 and 2 we display the results for the 
dispersive contributions stemming from $|\triangle J|=1$ and 
$|\triangle J|=0$ transitions\footnote{As customary, 
throughout this paper all the baryon polarizabilities are expressed 
in units of $10^{-4}\ {\rm fm}^3$.}. Of course, the total 
dispersive magnetic polarizability is the sum of these two pieces.
In these tables ``1st" indicates that only that intermediate state, 
which has the lowest excitation energy, is included while 
``1st + 2nd" refers to the sum (\ref{betad}) being cut after the 
next--to--lowest state. The total contribution is obtained by 
including all the intermediate states with an excitation energy
smaller than 3 GeV. Let us first discuss the $|\triangle J|=1$
contributions. Here we also consider the transition $\Lambda-\Sigma_0$
although both particles have $J=\frac{1}{2}$, because these two
particles are distinct by physical (isospin) quantum numbers rather
than orthogonal mixtures of higher $SU(3)$ representations. 
Otherwise the ``1st" state indeed 
corresponds to the observed $J=\frac{3}{2}$ baryon resonance which 
carries the same electrical and strangeness charges as the 
$J=\frac{1}{2}$ baryon under 
consideration. All other states (``2nd" and higher) are associated 
with higher $SU(3)$ excited $J=\frac{3}{2}$ states. We find that for 
all channels, which have a significant contribution from the ``1st"
state, (say, greater than one), the share carried by the excited 
states is almost negligible (less than 3\%). Only when the ``1st" 
contribution is small for some reason ({\it e.g.} it is U--spin 
forbidden as in the $\Sigma_-$ case \cite{Li73}) the "2nd" 
transition becomes important. In these channels the total dispersive 
magnetic polarizability nevertheless remains small. Basically, for 
all transitions the contributions from states higher than ``2nd" 
are negligible (less than 0.5\%). The RRA apparently exhibits only 
moderate deviations from the $SU(3)$ symmetry relations
\begin{eqnarray}
\beta_d(N-\Delta)=\beta_d(\Sigma_+-\Sigma_+^*)
=\beta_d(\Xi_0-\Xi_0^*)=\frac{4}{3}\beta_d(\Lambda-\Sigma_0^*) \, ,
\label{su3rel}
\end{eqnarray}
which are obtained by considering only the lowest intermediate 
state in Eq. (\ref{betad}). The SRA violates these relations by as 
much as 50 \%. Such a pattern has also been found for various other 
baryon properties \cite{We96}. From table 2 we observe that, as 
expected, the $|\triangle J|=0$ contributions are generally quite 
small. Again they are only recognizable when the corresponding 
$|\triangle J|=1$ transition is U--spin forbidden. Also, the 
contribution from the ``2nd" states is important only in some 
particular cases ({\it e.g.} $p$, $\Sigma_+$) while all 
contributions from states higher than ``2nd" may be discarded.

The total dispersive as well as the seagull contributions are given
in Table 3. There we not only compare the RRA and SRA but also quote
the results from the BSA \cite{GSS96}. For the total dispersive
part we see that the deviations from the symmetry relations
(\ref{su3rel}) in the BSA and RRA are opposite with respect to those 
of the SRA. This result is not completely unexpected because the 
first two approaches inaccurately predict the magnetic moment of the 
$\Sigma_+$ to be slightly larger or approximately equal to that of the 
proton \cite{Ku89,We96}, (in that case the symmetry relation in question
would read $\mu(\Sigma_+)=\mu(p)$ \cite{Ad85}). As mentioned above,
a major success of the SRA is the correct prediction of the pattern
of the magnetic moments, especially $\mu(\Sigma_+)/\mu(p)\approx0.85$
\cite{SW92}. For the seagull contributions we again recognize that 
the SRA yields sizable deviations from the symmetry relations
\begin{eqnarray}
\beta_s(p)=\beta_s(\Sigma_+)\qquad
\beta_s(n)=\beta_s(\Xi_0)\qquad
\beta_s(\Sigma_-)=\beta_s(\Xi_-)
\label{su3rels}
\end{eqnarray}
while neither the RRA nor the BSA do so. In case of the SRA 
these deviations cause $\beta_s$ to vary almost linearly with the 
strangeness charge, while the results from both RRA and BSA 
are roughly independent of strangeness. It is also somewhat 
surprising that while for the non--strange baryons ($p,n$) 
the predictions on $\beta_d$ are comparable in the RRA and 
SRA they differ by a factor two in case of $\beta_s$. This 
indicates that strange degrees of freedom play a significant
role inside the nucleon since in the infinite symmetry 
breaking limit, when the strange quarks are frozen out,
these two approaches yield identical -- $SU(2)$ -- results.

Up to now we have discussed the individual contributions 
separately. However, the physically relevant quantity rather is 
the total polarizability $\beta=\beta_d+\beta_s$. The 
corresponding predictions for $\beta$ are also given in 
Table 3. We observe that for the nucleon the SRA prediction 
is quite good because experiments favor a small positive number. 
The latest value quoted by the PDG \cite{PDG} is
$\beta(p)=2.1\pm0.8\pm0.5$. Comparison with the prediction of the 
non--relativistic quark model \cite{LM92} for $\beta(\Sigma_+)=1.7$
and $\beta(\Sigma_-)=-1.7$ also favors the SRA. 
However, as can be seen from table 3, any treatment of the 
three flavor Skyrme model leads to sizable isoscalar and 
isotensor contributions for the magnetic polarizabilities
in the $\Sigma$ channel.
In the $\Sigma_0$ channel the dispersive part is negative because 
this state dominantly couples to $\Lambda$ which has a lower mass. 
Hence this channel is the only one where dispersive and seagull 
parts add coherently indicating that the $\Sigma_0$ has the 
largest (in magnitude) magnetic polarizability. 

For completeness we also display the results for the electric seagull 
polarizability. The pertinent choice for the electromagnetic field is 
$A_\mu=(-\mbox{\boldmath $E$}\cdot\mbox{\boldmath $r$},
\bold{0})$.  
In the electric case the seagull contribution 
is  a good approximation to the total 
polarizability \cite{GSS96,Ch87,SC96}. 
In the collective treatment it is obtained from the matrix element
\cite{Sch93}
\begin{eqnarray}
\alpha^B_s &=&  \langle B \vert \left[
\gamma_\pi^{(e)}  D_{e,i}^2 + \gamma_K^{(e)} D_{e,\alpha}^2
\right]
\vert B \rangle \ .
\label{alphas}
\end{eqnarray}
Again, these matrix elements are evaluated in the space
of the collective coordinates. Furthermore 
\begin{eqnarray}
\gamma_\pi^{(e)} &=& {2e^2\over9} \int d^3 r \ r^2 s^2
 \left[ f_\pi^2 + {1\over{\epsilon^2}} (F'^2 + {s^2\over{r^2}}) +
 {2\over3} \ (f_K^2 - f_\pi^2)\ c \ \left( 1- D_{8,8} \right) \right]
\hskip1.5cm~
\\
\gamma_K^{(e)} &=& {e^2\over{6}} \int d^3 r \ r^2 (1-c)
\left[ f_K^2 + {1\over{4\epsilon^2}} (F'^2 +2{s^2\over{r^2}}) +
(f_K^2 - f_\pi^2) { c - 2 \over3}
\left( 1- D_{8,8} \right) \right] .
\end{eqnarray}
Here we have omitted non--minimal photon couplings since,
practically, they give no contribution to the electric
polarizabilities (see footnote 3 in ref \cite{GSS96} for details 
on this issue).  In table 4 the numerical results are compared to 
the corresponding predictions of the BSA. We observe that the SRA 
prediction of the electric seagull polarizability for the nucleon 
($\alpha_s(p)=11.2,\alpha_s(n)=11.0$) agrees reasonably well with 
the PDG data: $\alpha(p)=12.1\pm0.8\pm0.5$ and
$\alpha(n)=9.8^{+1.9}_{-2.3}$.
Both, the RRA and the BSA yield numbers which are about twice as
large. As the collective structures of the operators in 
(\ref{alphas}) and (\ref{betas}) are identical not only the 
relations analogous to (\ref{su3rels}) hold in the symmetric 
case but also the above discussed deviations from the flavor 
symmetric predictions are similar for the electric and magnetic 
seagull contributions. To a good accuracy the seagull pieces obey 
$\alpha_s=-2\beta_s$ in all three treatments. This implies that the 
Skyrme term is only of minor importance.

We have seen that various treatments of the three flavor 
generalization of the Skyrme model yield quite different 
results for the electromagnetic polarizabilities. In particular 
the deviations from the $SU(3)$ symmetry relations for the 
dispersive parts are quite different while at least the 
seagull parts in the BSA and RRA are quite similar. Actually 
similarities between the BSA and RRA are expected from the 
computation of many other observables \cite{We96,Sch95}. Comparing 
especially the symmetry breaking pattern for the predicted magnetic 
moments of the $\frac{1}{2}^+$ baryons with the experimental data 
however favors the SRA. Available data on the 
nucleon polarizabilities tend to support this assessment. It is thus 
suggestive that the pattern of the electromagnetic polarizabilities of 
the low--lying $\frac{1}{2}^+$ baryons should follow the predictions of 
the SRA to the $SU(3)$ Skyrme model. 

Let us finally add a word of caution 
concerning the quantitative results. As is well--known, neither
of the three approaches discussed here correctly predicts 
the absolute values of the baryon magnetic moments, {\it e.g.}
the magnetic moment of the proton is found to be (in nucleon
magnetons) $1.77$, $1.68$ and $1.78$ in the bound state, the 
rigid rotator and the slow rotator approaches, respectively.
This is to be compared with the actual value of $2.79$. This 
insufficiency is inherited from the $SU(2)$ Skyrme model, but it is 
also cured there. A recent study has shown that the moments at 
${\cal O}(N_C)$ plus the quantum corrections at next to 
leading order, ${\cal O}(N_C^0)$, fill the gap \cite{Me96}. General 
considerations of the $1/N_c$--expansion show that the magnetic 
moment operator $\mu_3$ acquires a multiplicative correction 
\cite{Da94}. Since this operator crucially enters the dispersive 
parts of the magnetic polarizabilities (\ref{betad}) a change in 
the numerical results would not be unexpected. Similar corrections
may also arise for the seagull component of the magnetic 
polarizabilities. The computation of the electric polarizabilities 
in the two flavor models also shows that loop corrections 
to the corresponding ${\cal O}(N_C)$ seagull components 
are important\cite{Me96}. Whether this statement carries over 
to $SU(3)$ remains subject to further studies.
 
\bigskip

Prof. B. Schwesinger passed away shortly after this article was
submitted for publication. HW and NNS would like to express all their
gratitude to him as teacher, collegue and friend.

\pagebreak
 
\begin{table}
\begin{center}
Table 1: Dispersive contributions $\beta_d$ to the static 
magnetic polarizabilities corresponding to transitions between
different baryons (mostly $|\triangle J|=1$ transitions). 
All data are in $10^{-4} {\rm fm}^3$.
\vspace{1.cm}

\begin{tabular}{|c||c|c|c|c|c|c|} \hline
  & \multicolumn{3}{|c|}{RRA} & \multicolumn{3}{|c|}{SRA} \\ \hline  
      & 1st & 1st + 2nd & Total & 1st & 1st + 2nd & Total \\  \hline
${N - \Delta}$ 
& 4.529 & 4.559 & 4.559 & 5.621 & 5.703 & 5.709 \\ 
\hline 
${\Lambda - \Sigma_0}$ 
& 4.031 & 4.093 & 4.098 & 3.479 & 3.581 & 3.586 \\
\hline 
${\Lambda - \Sigma_0^*} $ 
& 3.835 & 3.875 & 3.877 & 3.237 & 3.318 & 3.322 \\
\hline
${\Sigma_+ - \Sigma_+^*}$ 
& 4.954 & 5.200 & 5.204 & 3.512 & 3.600 & 3.605 \\ 
\hline 
${\Sigma_0 - \Sigma_0^*}$ 
& 0.875 & 1.067 & 1.070 & 0.572 & 0.657 & 0.659 \\
\hline 
${\Sigma_- - \Sigma_-^*}$ 
& 0.126 & 0.270 & 0.275 & 0.130 & 0.213 & 0.214 \\ 
\hline
${\Xi_0 - \Xi_0^*}$ 
& 5.060 & 5.419 & 5.423 & 2.873 & 2.952 & 2.956 \\ 
\hline
${\Xi_- - \Xi_-^*}$
& 0.134 & 0.503 & 0.504 & 0.062 & 0.224 & 0.225 \\ 
\hline   
\end{tabular}
\end{center}
\end{table}


\begin{table}
\begin{center}
Table 2: Dispersive contributions $\beta_d$ to the static 
magnetic polarizabilities corresponding to the $|\triangle J|=0$
transitions. The superscript $exc$ refers to $SU(3)$ excited 
states. All data are in $10^{-4} {\rm fm}^3$.
\vspace{1.cm}

\begin{tabular}{|c||c|c|c|c|c|c|} \hline
  & \multicolumn{3}{|c|}{RRA} & \multicolumn{3}{|c|}{SRA} \\ \hline  
      & 1st & 1st + 2nd & Total & 1st & 1st + 2nd & Total \\  \hline
${p - p^{exc}}$ 
& 0.010 & 0.042 & 0.042 & 0.014 & 0.051 & 0.051 \\ 
\hline
${n - n^{exc}}$ 
& 0.081 & 0.085 & 0.085 & 0.098 & 0.098 & 0.100 \\  
\hline
${\Lambda - \Lambda^{exc}}$ 
& 0.017 & 0.017 & 0.017 & 0.051 & 0.051 & 0.051 \\
\hline 
${\Sigma_+ - \Sigma_+^{exc}}$ 
& 0.023 & 0.079 & 0.080 & 0.005 & 0.030 & 0.030 \\ 
\hline 
${\Sigma_0 - \Sigma_0^{exc}}$ 
& 0.086 & 0.110 & 0.111 & 0.021 & 0.033 & 0.033 \\
\hline 
${\Sigma_- - \Sigma_-^{exc}}$ 
& 0.174 & 0.185 & 0.186 & 0.130 & 0.133 & 0.134 \\ 
\hline
${\Xi_0 - \Xi_0^{exc}}$ 
& 0.011 & 0.011 & 0.011 & 0.0003 & 0.0003 & 0.0003 \\ 
\hline
${\Xi_- - \Xi_-^{exc}}$
& 0.029 & 0.029 & 0.029 & 0.008 & 0.008 & 0.008 \\ 
\hline   
\end{tabular}
\end{center}
\end{table}

\begin{table}
\begin{center}
Table 3: Total magnetic polarizabilities as the sum of the 
dispersive and seagull contributions, {\it i.e.}
$\beta=\beta_d+\beta_s$. Results are listed according to 
different approaches to the SU(3) Skyrme model, see text.
All data are in $10^{-4} {\rm fm}^3$.
\vspace{1.cm}

\begin{tabular}{|c||c|c|c|c|c|c|c|c|c|} \hline
  &\multicolumn{3}{|c|}{BSA \protect\cite{GSS96}}
  &\multicolumn{3}{|c|}{RRA} 
  &\multicolumn{3}{|c|}{SRA} \\  \hline
  & $\beta_d$ & $\beta_s$ & $\beta$ 
  & $\beta_d$ & $\beta_s$ & $\beta$ 
  & $\beta_d$ & $\beta_s$ & $\beta$ \\ \hline
$p$  &
5.6  & --  8.3  & -- 2.7  &
4.6 & -- 10.2 & -- 5.6 &
5.8 & --  5.3 &    0.5 \\ 
\hline 
$n$  & 
5.6  & -- 8.3  & -- 2.7  &
4.6 & --10.0 & -- 5.3 &
5.8 & -- 5.2 &    0.6 \\ 
\hline 
$\Lambda$ &
12.1  & -- 8.7  &   3.4  &
8.0  & -- 9.8 & --1.8 & 
7.0  & -- 3.2 &   3.6 \\
\hline 
$\Sigma_+$ & 
10.4  & --  9.1  &   1.3    &
5.3   & -- 10.7 & --5.4   &  
3.6   & -- 3.1  &   0.5  \\  
\hline 
$\Sigma_0$ &
-- 4.0   & -- 8.7   & -- 12.7   &  
-- 2.7  & -- 9.6  & -- 12.3   &  
-- 2.9  & -- 2.8  & --  5.6   \\
\hline 
$\Sigma_-$ &
0.5   & -- 8.4   & -- 7.9   &  
0.5   & -- 8.4  & -- 8.0    &  
0.4   & -- 2.4  & -- 2.1    \\ 
\hline
$\Xi_0$    &
14.0   & -- 9.6   &    4.4   &  
 5.4   & --10.4  & -- 5.0   &  
 3.0   & -- 2.3  &    0.7    \\ 
\hline
$\Xi_-$    &
1.5   & -- 8.7   & -- 7.2   &  
0.5   & -- 7.7  & -- 7.2   &  
0.2   & -- 1.7  & -- 1.4    \\ 
\hline   
\end{tabular}
\end{center}
\end{table}
\begin{table}
\begin{center}
Table 4: The electric polarizabilities as approximated by their 
seagull contributions (\protect\ref{alphas}) in various treatments 
of the SU(3) Skyrme model. All data are in $10^{-4}{\rm fm}^3$.
\vspace{1.cm}

\begin{tabular}{|c||c|c|c|}\hline
&BSA \protect\cite{GSS96}&RRA&SRA \\ \hline
$p$  &
17.3 &  20.9  & 11.2 \\ \hline
$n$  &
17.3 &  20.5  & 11.0 \\ \hline
$\Lambda$ &
18.1  & 20.1  & 7.0 \\ \hline
$\Sigma_+$ &
18.1 & 22.0 &  6.6 \\ \hline
$\Sigma_0$ &
18.8 & 19.7 & 5.9 \\ \hline
$\Sigma_-$ &
17.4  & 17.3  & 5.1 \\ \hline
$\Xi_0$    &
19.9   & 21.3 & 4.9 \\ \hline
$\Xi_-$    &
18.0   & 15.7 & 3.5 \\ \hline
\end{tabular}
\end{center}
\end{table}

\end{document}